\documentclass[aps,prl,shownopacs,superscriptaddress,twocolumn]{revtex4}

\usepackage{graphicx}
\usepackage{hyperref}
\usepackage{amsmath}
\usepackage{amssymb}
\usepackage{stmaryrd}
\usepackage{epstopdf}
\usepackage{bbm}
\usepackage{pgfplotstable}
\usepackage{physics}
\usepackage{bbold}
\usepackage{gensymb}
\usepackage{siunitx}

\begin{document}

\title{Demonstration of Einstein-Podolsky-Rosen Steering \\Using Hybrid Continuous- and Discrete-Variable Entanglement of Light}

\author{A. Cavaill\`{e}s}
\affiliation{Laboratoire Kastler Brossel, Sorbonne Universit\'e, CNRS, ENS-Universit\'e PSL, Coll\`ege de France, 4 Place
Jussieu, 75005 Paris, France}
\author{H. Le Jeannic\footnotemark[2]\footnotetext{\footnotemark[2] Present address: Niels Bohr Institute, University of Copenhagen, Blegdamsvej 17, DK-2100 Copenhagen, Denmark.}}
\affiliation{Laboratoire Kastler Brossel, Sorbonne Universit\'e, CNRS, ENS-Universit\'e PSL, Coll\`ege de France, 4 Place
Jussieu, 75005 Paris, France}
\author{J. Raskop}
\affiliation{Laboratoire Kastler Brossel, Sorbonne Universit\'e, CNRS, ENS-Universit\'e PSL, Coll\`ege de France, 4 Place
Jussieu, 75005 Paris, France}
\author{G. Guccione}
\affiliation{Laboratoire Kastler Brossel, Sorbonne Universit\'e, CNRS, ENS-Universit\'e PSL, Coll\`ege de France, 4 Place
Jussieu, 75005 Paris, France}
\author{D. Markham}
\affiliation{Laboratoire d'Informatique de Paris 6, Sorbonne Universit\'e, CNRS, 4 Place Jussieu, 75005 Paris, France}
\author{E. Diamanti}
\affiliation{Laboratoire d'Informatique de Paris 6, Sorbonne Universit\'e, CNRS, 4 Place Jussieu, 75005 Paris, France}
\author{M. D. Shaw}
\affiliation{Jet Propulsion Laboratory, California Institute of Technology, 4800 Oak Grove Drive, Pasadena, California 91109, USA}
\author{V. B. Verma}
\affiliation{National Institute of Standards and Technology, 325 Broadway, Boulder, CO 80305, USA}
\author{S. W. Nam}
\affiliation{National Institute of Standards and Technology, 325 Broadway, Boulder, CO 80305, USA}
\author{J. Laurat}
\email{julien.laurat@sorbonne-universite.fr}
\affiliation{Laboratoire Kastler Brossel, Sorbonne Universit\'e, CNRS, ENS-Universit\'e PSL, Coll\`ege de France, 4 Place
Jussieu, 75005 Paris, France}

\date{\today}

\begin{abstract}
Einstein-Podolsky-Rosen steering is known to be a key resource for one-sided device-independent quantum information protocols. Here we demonstrate steering using hybrid entanglement between continuous- and discrete-variable optical qubits. To this end, we report on suitable steering inequalities and detail the implementation and requirements for this demonstration. Steering is experimentally certified by observing a violation by more than $5$ standard deviations. Our results illustrate the potential of optical hybrid entanglement for applications in heterogeneous quantum networks that would interconnect disparate physical platforms and encodings.
\end{abstract}

\maketitle

The capacity for one part of a two-party entangled quantum system to steer the other through measurement is a fundamental feature of quantum mechanics and has recently been recognized as a useful resource for quantum networks \cite{cavalcanti_quantum_2017}. This capacity, known as Einstein-Podolsky-Rosen steering (EPR steering), lies between entanglement and Bell-non locality~\cite{wiseman_steering_2007,brunner_bell_2014}. EPR steering has applications in various protocols when only one party can be trusted, called one-sided device independent, including quantum key distribution ~\cite{branciard_one-sided_2012}, randomness generation~\cite{law_quantum_2014, passaro_optimal_2015}, entanglement estimation~\cite{Toth+2015EntSteering}, and entanglement verification for quantum networks~\cite{Cavalcalnti + Ass q networks,McCutcheon}. It can also be used to test underlying properties of systems and measurements in the form of self-testing and rigidity statements, with broad applications for example in delegated quantum computing~\cite{Gheorghiu+Rigidity steering delegated qc,SupicHoban16 self testing through steering}.

Such strong motivations led to a variety of steering demonstrations. These pioneering works followed the traditional separation in quantum information science between discrete- (DV) and continuous-variable (CV) approaches. They were realized with light either in discrete-variable systems verifying polarization entanglement~\cite{saunders_experimental_2010,smith_conclusive_2012,wittmann_loophole-free_2012,weston_heralded_2018} or path entanglement for single photons~\cite{fuwa_experimental_2015,guerreiro_demonstration_2016}, or in continuous-variable systems using Gaussian states~\cite{Handchen2012,Armstrong2015,walk_experimental_2016,Deng2017,Qin2017}. However, in recent years, a growing body of works appeared to bridge the two CV and DV approaches in single \textit{hybrid} experiments \cite{van_loock_optical_2011,andersen_hybrid_2015}. Experimental realizations of hybrid protocols include for instance the teleportation of DV quantum bits using a continuous-variable teleporter~\cite{takeda_deterministic_2013} or the development of a CV witness for single-photon entanglement~\cite{morin_witnessing_2013,ho_witnessing_2014}. The recent demonstration of hybrid entanglement of light between discrete- and continuous-variable qubits~\cite{jeong_generation_2014,morin_remote_2014} paved the way towards the realization of hybrid quantum networks allowing for the transfer of information between CV and DV nodes. Such networks require however the implementation of robust entanglement verification schemes to perform either fully or one-sided device-independent protocols. 

In this Letter, we report on the first demonstration of EPR steering using a hybrid CV-DV entangled state. A steering test free of post-selection is implemented using high-efficiency homodyne detections. Steering is then conclusively certified through quantum tomography and using semi-definite programming. We also provide detailed characterization of the requirements to achieve this task. This demonstration realized in a scenario amenable to one-sided device-independent schemes is an important step towards operational protocols in heterogeneous quantum networks.

\textit{Principle.}--- In a steering scenario, Alice and Bob share an entangled state and Alice, who cannot be trusted, has to convince Bob that she can remotely steer his system. To this end she performs a measurement $\theta$, which yields a result $a$. This information is then sent to Bob. Depending on the measurement and result, Bob's system is projected to the state $\rho_{a|\theta}$ with probability $p(a|\theta)$. Whether steering is observed or not will be determined by the information contained in the set $\{p(a|\theta), \rho_{a|\theta}\}_{a,\theta}$ obtained after repeated measurements. Equivalently, one could consider the set of unnormalized states $\{\sigma_{a|\theta}\}_{a,\theta}$, called assemblage, defined by $\sigma_{a|\theta}=p(a|\theta)\rho_{a|\theta}$. 

In our case, Alice and Bob share the hybrid entangled optical state $\ket{\Psi}_\textrm{AB}$ initially demonstrated in \cite{morin_remote_2014}
\begin{equation}
\ket{\Psi}_\textrm{AB}=\sqrt{R}\ket{0}_\textrm{A}\!\ket{\textrm{CSS}_{-}}_\textrm{B}-\sqrt{1-R}\ket{1}_\textrm{A}\!\ket{\textrm{CSS}_{+}}_\textrm{B}.
\end{equation}
The vacuum and single-photon states $\ket{0}$ and $\ket{1}$ form the basis of Alice's DV mode, while Bob's CV mode is populated by the coherent-state superpositions $\ket{\textrm{CSS}_{\pm}}\propto(\ket{\alpha}\pm\ket{-\alpha})$, which are also known as optical ``Schr\"{o}dinger's cat'' states. 

\begin{figure}[t!]
\centering
\includegraphics[width=0.93\columnwidth]{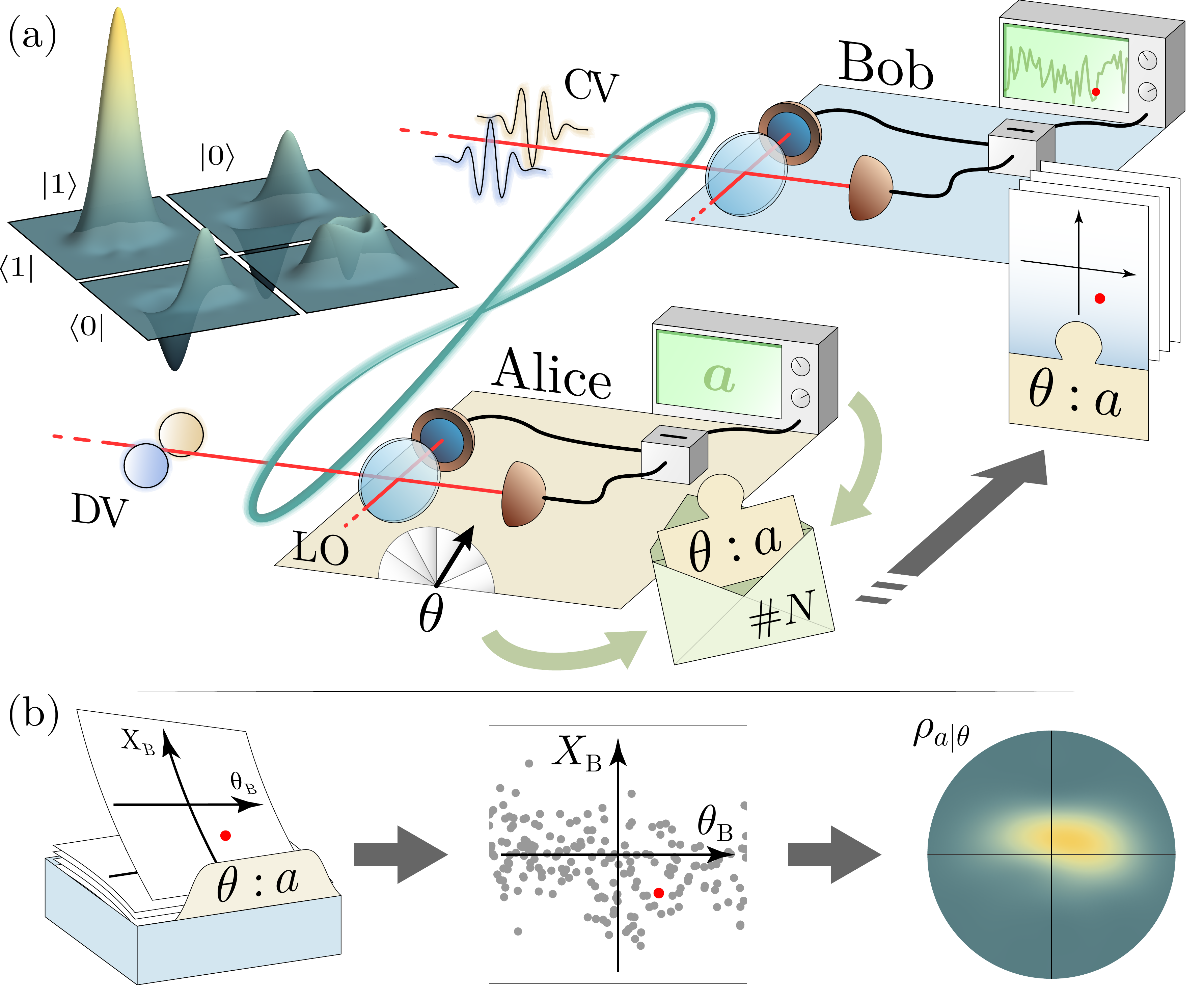}
\caption{Steering scenario with hybrid CV-DV entanglement of light. (a) The two-mode hybrid entangled state is shared between Alice, who cannot be trusted, and Bob. On the DV mode, Alice locally performs quadrature measurements using homodyne detection at different phases $\theta$ of her local oscillator (LO) and registers the sign-binned measurement result $a=\pm$. She then sends the information to Bob, who uses it to sort his own quadrature measurements depending on the phase choice and sign result $\{\theta, a\}$. (b) Via quantum state tomography, Bob is able to reconstruct each conditional state $\rho_{a|\theta}$ and the associated Wigner functions. As detailed in the text, he obtains the assemblage $\{\sigma_{a|\theta}\}_{a,\theta}$ and tests it against any LHS model to prove that EPR steering has occurred.}
\label{fig1}
\end{figure}

The steering scenario is detailed in Fig. \ref{fig1}. Both parties use a homodyne detection setup, in Alice's case to perform quadrature measurements at phase $\theta$ and in Bob's case to perform tomographic reconstruction of the states $\rho_{a|\theta}$. For each heralding event announcing the entangled state generation (labelled by $\#N$), Alice chooses a measurement by tuning the relative phase of her local oscillator $\theta$ among a set of $m_\textrm{A}$ possible choices, then registers the result $a$ obtained by sign binning, \text{i.e.}, separates them into two possible outcomes $\{+,-\}$. A quadrature measurement $q_\textrm{A}$ along the relative phase of the local oscillator $\theta$ remotely prepares Bob's system into the state \cite{RSP}
\begin{equation}
\bra{q_\textrm{A}}\ket{\Psi}_\textrm{AB}\propto\sqrt{R}\ket{\textrm{CSS}_{-}}_\textrm{B}-q_\textrm{A}e^{i\theta}\sqrt{1-R}\ket{\textrm{CSS}_{+}}_\textrm{B}.
\end{equation}
This results in a set of $2m_\textrm{A}$ conditional states evenly spread throughout the phase space (see Supplemental Material \cite{SM}):
\begin{multline}
\sigma_{a|\theta}=p(a|\theta)\big(R\ket{\textrm{CSS}_{-}}\!\bra{\textrm{CSS}_{-}} + (1-R)\ket{\textrm{CSS}_{+}}\!\bra{\textrm{CSS}_{+}}\\+a\,e^{i\theta}\sqrt{2R(1-R)/\pi}\ket{\textrm{CSS}_{-}}\bra{\textrm{CSS}_{+}}+\textrm{h.c.}\big),
\end{multline}
where $p(a|\theta)=1/2$ for all $a$ and $\theta$ as positive and negative quadrature measurements are equally likely.

Bob separately performs tomographic measurements by recording the relative phase of his local oscillator and the quadrature value obtained at each event $\#N$. Having no information from Alice for this particular event, Bob cannot see the effect of her measurements yet. It is only after associating Alice's observations $\{\theta, a\}_N$ to the corresponding heralding events that he is able to split his data into $2m_\textrm{A}$ subsets. He can then reconstruct the states $\{\rho_{a|\theta}\}_{a,\theta}$ and obtain the assemblage $\{\sigma_{a|\theta}\}_{a,\theta}$. EPR steering is demonstrated if the assemblage is impossible to describe with a \textit{local hidden state} (LHS) model~\cite{wiseman_steering_2007}, in which case Bob will be convinced and the test successful.


\textit{Semidefinite programming and steering inequalities.}--- A LHS strategy consists of Bob receiving a \textit{local} quantum state $\rho_\lambda$ while Alice receives a related piece of classical information $\lambda$ that will determine the result $a$ of her measurement $\theta$ according to a probability $p(a|\theta,\lambda)$. Denoting by $\mu(\lambda)$ the distribution of all $\lambda$, an assemblage $\{\sigma_{a|\theta}\}$ following an LHS model satisfies
\begin{equation}
\sigma_{a|\theta}= \int d\lambda\mu(\lambda)p(a|\theta,\lambda)\rho_{\lambda}\indent\forall\,a,\theta.
\label{LHS}
\end{equation}
Checking an assemblage against any LHS representation is hard in the general case~\cite{cavalcanti_quantum_2017}, but it is possible to simplify the problem when the number of measurements and outputs made by Alice is finite. In that case one can indeed reformulate the task of verifying \eqref{LHS} as a semidefinite program (SDP)~\cite{vandenberghe_semidefinite_1996}, \textit{i.e.}, a convex optimization problem that can be solved efficiently. 

In the framework of the SDPs we will consider, the condition for an existing LHS model can be written in the form of \textit{steering inequalities}: For any set of suitable~\cite{cavalcanti_quantum_2017} operators $\{F_{a|\theta}\}_{a,\theta}$, if assemblage $\{\sigma_{a|\theta}\}$ has an LHS model, then
\begin{equation}
\mathcal{S}\hat{{}={}}\operatorname{Tr}\Big(\sum_{a,\theta}F_{a|\theta}\sigma_{a|\theta}\Big)\geq0.
\label{SF}
\end{equation}
The SDP aims to minimise $\mathcal{S}$ over all valid $\{F_{a|\theta}\}_{a,\theta}$. If the minimum computed value $\mathcal{S}\textsubscript{min}$ is negative, the LHS model is not fit to describe the experiment, thereby demonstrating EPR steering. In the following we will use the set of operators that provides $\mathcal{S}\textsubscript{min}$ to define the optimal steering inequality. 


\textit{Experimental implementation.}--- The hybrid entangled state is generated following the measurement-induced method demonstrated in~\cite{morin_remote_2014} that enables to create it at a distance, even with a very lossy channel between the two parties. It relies on two optical parametric oscillators (OPO), one located with each party, pumped below threshold by a continuous-wave frequency-doubled Nd:YAG laser. The first OPO, based on a type-II phase-matched KTP crystal, is used as the DV source and enables on its own the generation of single photons, with a heralding efficiency above 90\%, using high-efficiency superconducting nanowire single-photon detectors~\cite{jeannic_high-efficiency_2016}. The second OPO is based on a PPKTP crystal and generates a 3-dB-squeezed vacuum state. For $\lvert\alpha\rvert^2\sim1$ this state presents close-to-unity fidelity with $\ket{\textrm{CSS}_{+}}$, as does the photon-subtracted squeezed vacuum state with $\ket{\textrm{CSS}_{-}}$ in the same conditions. This latter state can be heralded by tapping off a small amount of power at the output of the OPO and detecting a single photon. The entangled state $\ket{\Psi}_\textrm{AB}$ is generated by mixing in an indistinguishable fashion the two heralding paths. The parameter $R$ can be varied by adjusting the ratio between the heralding rates of each separate source. 

In order to model the experiment, several imperfections have to be taken into account. First, one needs to consider the overall transmission losses on Alice and Bob's modes, respectively $\eta_\textrm{A}$ and $\eta_\textrm{B}$. The escape efficiency of both OPOs is estimated to be \SI{90}{\percent}~\cite{morin_quantum_2014}, which directly translates to \SI{10}{\percent} of intrinsic loss on both modes. Additionally, Alice's and Bob's measurements are performed using a homodyne setup that introduces overall \SI{15}{\percent} of detection loss. The asymmetry of the steering scenario implies however that these losses will not have the same impact for the two parties. Indeed, Bob performs a full tomography to reconstruct the assemblage using a Maximum-Likelihood (MaxLik) algorithm \citep{lvovsky_iterative_2004,morin_experimentally_2013} and is therefore able to correct for detection losses. This is acceptable as it is assumed Bob has full knowledge of his setup and of the losses introduced by his apparatus. On the other hand, no assumption can be made with regards to Alice's measurements, meaning that a similar correction process is not acceptable. The last imperfection relates to Alice's quadrature measurements, which are implemented by microcontroller-based locking~\cite{huang_microcontroller-based_2014} of her local oscillator phase. The associated noise exhibits a standard deviation of \SI{3}{\degree} over the acquisition.

\begin{figure}[t]
\centering
\includegraphics[width=0.9\columnwidth]{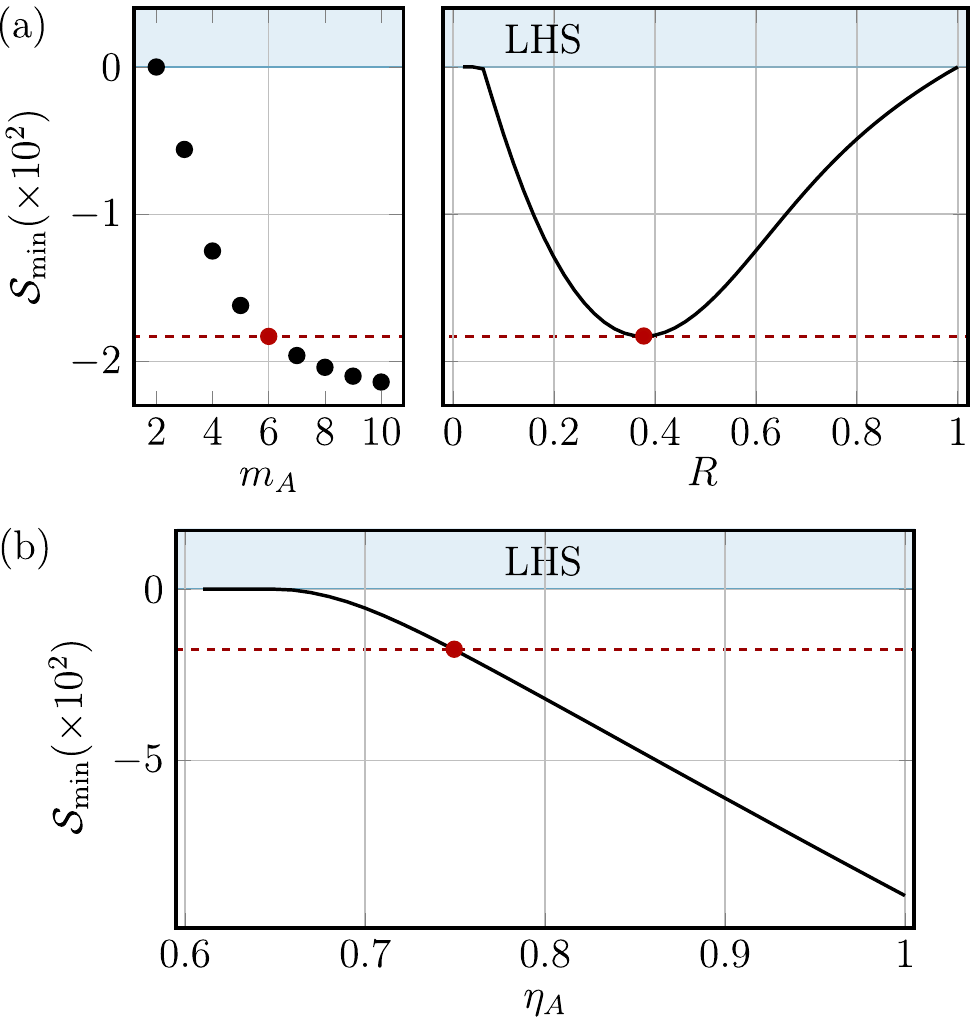}
\caption{Expected maximum steering inequality violation. (a) Left: maximum violation depending on the number of measurements $m_\textrm{A}$ performed by Alice, calculated at the optimal ratio $R$. Right: expected violation as a function of $R$ for $m_\textrm{A}=6$. In both panels the overall efficiencies are assumed to be $\eta_\textrm{B}=\SI{90}{\percent}$ for Bob and $\eta_\textrm{A}=\SI{75}{\percent}$ for Alice. (b) Maximum violation as a function of Alice's efficiency $\eta_\textrm{A}$ for $m_\textrm{A}=6$, $\eta_B=\SI{90}{\percent}$, and $R=0.37$. The red dot indicates the maximal violation possible in our experimental conditions.}
\label{fig2}
\end{figure}

\textit{Expected steering inequality violation.}--- Given these experimental imperfections, the possibility of observing a steering inequality violation was explored using SDP codes supplied in~\cite{cavalcanti_quantum_2017} to choose the best settings. Figure~\ref{fig2} provides the evolution of $\mathcal{S}_{\textrm{min}}$ with various experimental settings. 

The first setting choice is the number $m_\textrm{A}$ of measurements performed by Alice. The left panel of Fig.~\ref{fig2}$(a)$ shows the largest possible violation as a function of this number. In our experimental conditions, one can see that at least three measurements are needed; more measurements translate to a larger theoretical violation. As more measurements also complicate the experimental analysis, we chose to limit to six measurements on Alice's side. The variation of $\mathcal{S}_{\textrm{min}}$ as a function of $R$ for $m_\textrm{A}=6$ is then given on the right panel of Fig.~\ref{fig2}$(a)$. Because of the asymmetry in transmission losses for the CV and DV modes, equal balance between the two heralding rates is not optimal and the best violation is found for an unbalanced ratio $R=0.37$. 

For these parameters, Figure~\ref{fig2}$(b)$ finally presents the best violation attainable depending on the transmission efficiency on Alice's side, which is not possible to correct for. One can see that our demonstration of EPR steering is challenging since it requires an overall transmission efficiency higher than $\eta_\textrm{A}=\SI{65}{\percent}$. At our experimental value $\eta_\textrm{A}\approx\SI{75}{\percent}$ a steering violation can be expected.

\begin{figure}[t]
\centering
\includegraphics[width=0.85\columnwidth]{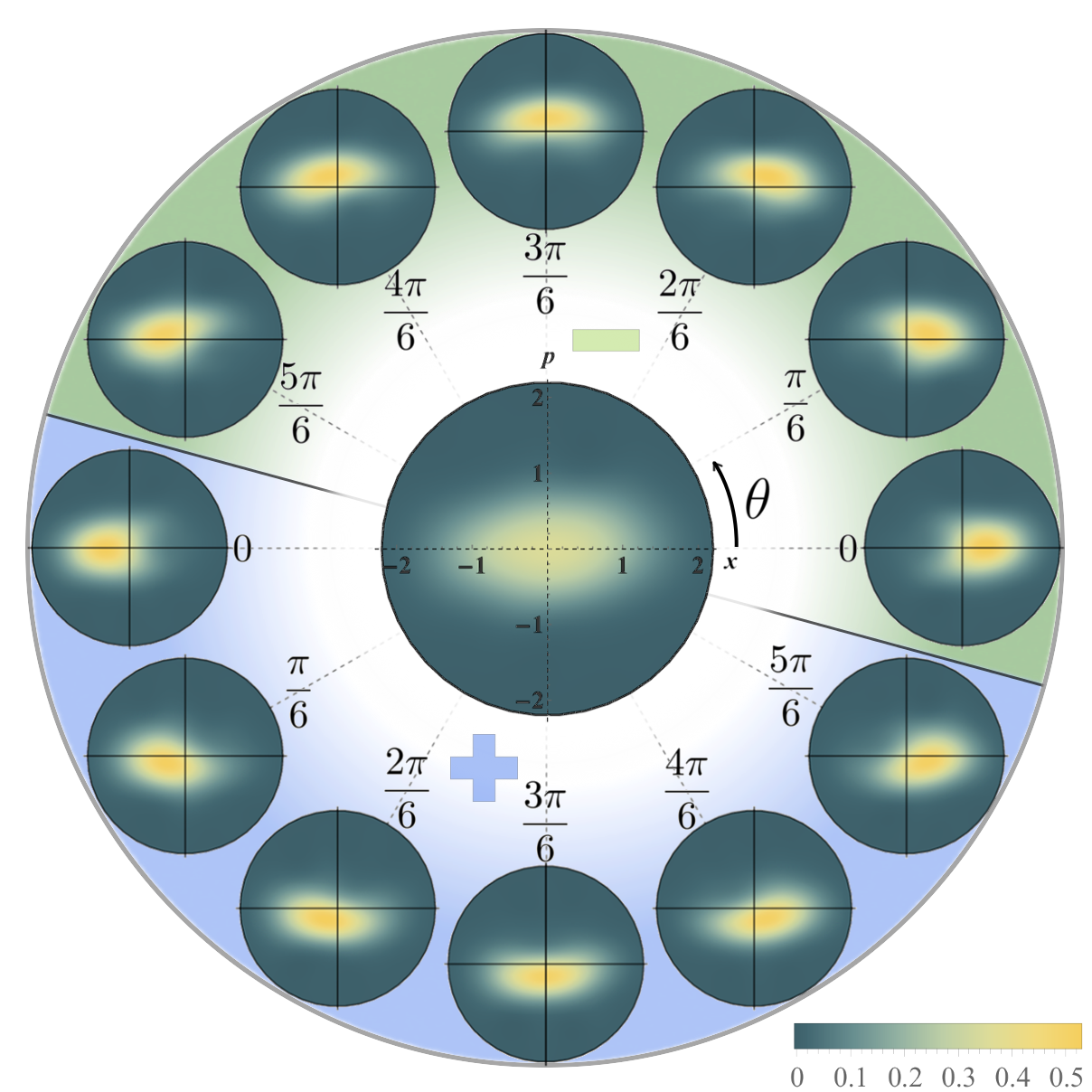}
\caption{Experimental steered states. The Wigner function of Bob's unconditioned state is shown at the center of the disk. When Alice makes measurement $\theta\in\{n.\frac{\pi}{6}\}$ and gets result $a~=~\pm$, Bob's state is projected to the corresponding $\rho_{a|\theta}$. The associated Wigner function is displayed at the angle $\theta$ and in the half-space corresponding to $a$. The steering resulting from Alice's measurements translates into a rotation of Bob's state around the phase-space origin.}
\label{fig3}
\end{figure}

\textit{Experimental steered states.}--- We generated the hybrid entangled state with $R=0.36$, in line with our simulation for optimal violation and within our experimental precision. The state, which was heralded at a rate of 200~kHz, was first checked using the homodyne setups available on both the CV and DV modes, and the two-mode Wigner function was reconstructed using a MaxLik algorithm. The single-mode Wigner functions of the four projections on the DV mode $\bra{i}\rho_{\textrm{AB}}\ket{j}$ with $i,j\in\{0,1\}$  are plotted on the left of Fig.~\ref{fig1}. Higher-photon-number components in the DV mode are limited to $2\%$. The entanglement negativity reaches $\mathcal{N}=0.28\pm0.01$ when corrected for detection losses \cite{morin_remote_2014,Vidal}.

After characterizing the hybrid state, we realized the steering test with $m_\textrm{A}=6$ measurements performed in sequence by Alice, corresponding to $6$ values of her LO's relative phase $\theta=n\times\frac{\pi}{6}$ with $n\in[0,5]$. An accurate reconstruction of the assemblage required the accumulation of $120 000$ quadrature measurements on Bob's side for each value $\theta$. Using a MaxLik algorithm, we were able to reconstruct the complete assemblage $\{\sigma_{a|\theta}\}_{a,\theta}$.

The experimental Wigner functions of each subset of the assemblage are presented in Fig.~\ref{fig3}. The unconditioned state is shown at the center, $\sigma_\theta=\sum_a\sigma_{a|\theta}$. The non-signaling condition $\sum_a\sigma_{a|\theta}=\sum_a\sigma_{a|\theta'}\indent$ for all $\theta$, $\theta'$ required for a valid steering test~\cite{cavalcanti_quantum_2017} is verified here, since we measure an average fidelity between unconditioned states $\mathcal{F}(\sigma_{\theta},\sigma_{\theta'})=99.7\pm0.1\%$, within the bounds of the typical uncertainties associated with a MaxLik reconstruction. The Wigner functions of the 12 conditional states are displayed along the perimeter of Fig.~\ref{fig3}, at the angle and in the half space respectively corresponding to Alice's choice $\theta$ and result $a=\pm$. As expected, depending on Alice's measurements, Bob's conditional state rotates around the phase-space origin.

\begin{figure}[t]
\centering
\includegraphics[width=0.92\columnwidth]{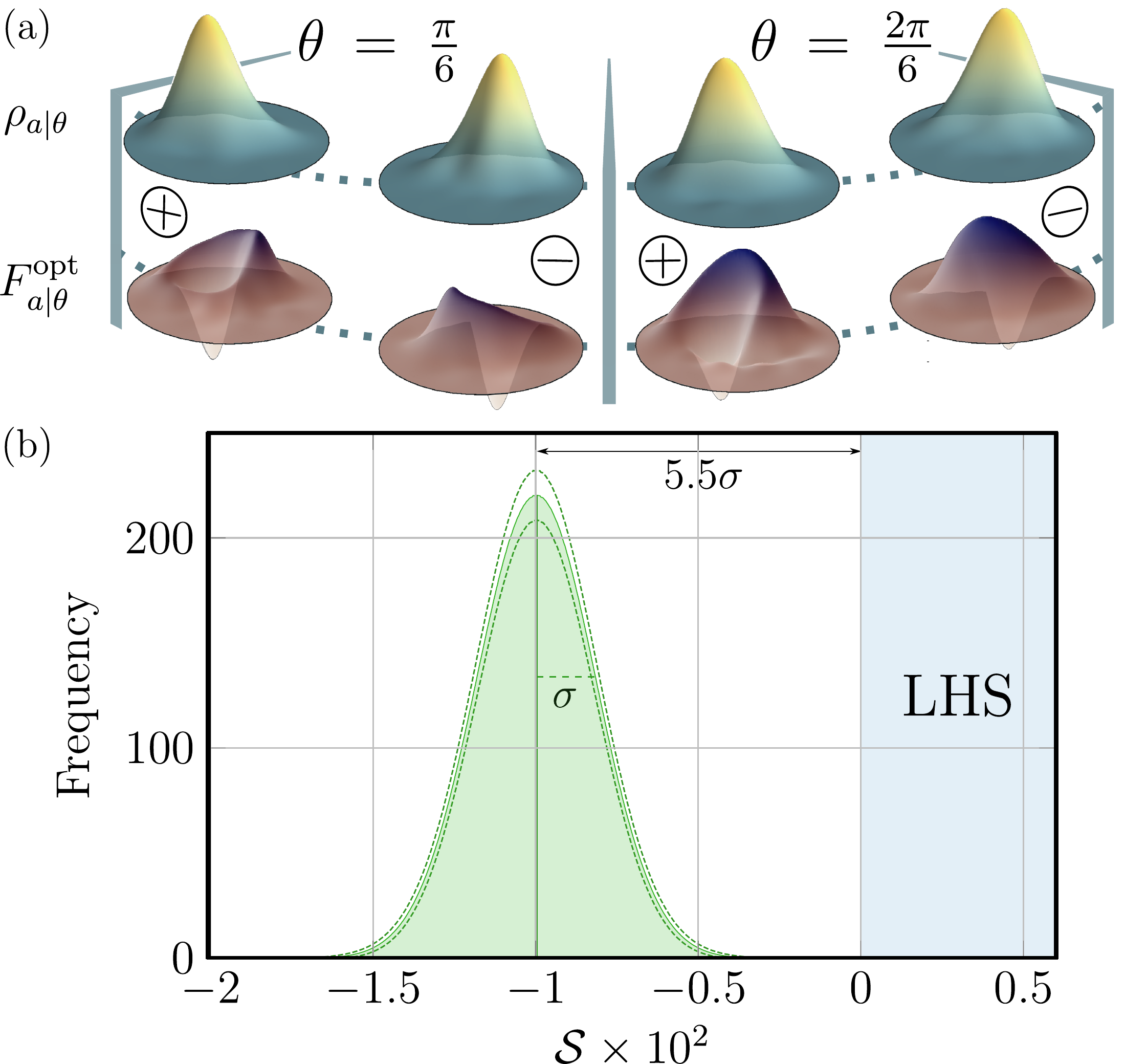}
\caption{Experimental violation. (a) Examples of Wigner functions of Bob's states $\rho_{a|\theta}$, conditioned here by Alice's measurements $\{\frac{\pi}{6},\frac{2\pi}{6}\}$ giving results $\{+,-\}$, and of the corresponding optimal steering inequality operators $F_{a|\theta}^{\textrm{opt}}$ determined by SDP. (b) Histogram of $\SI{5E7}{}$ values of $\mathcal{S}~=~\textrm{Tr}\left(\sum_{a,\theta}F_{a|\theta}^{\textrm{opt}}\sigma_{a|\theta}\right)$ taking into account the errors induced by the Maximum-Likelihood reconstruction via a Metropolis-Hastings algorithm. The dashed lines delimit the area of one standard deviation confidence on the distribution. The measured mean value is equal to $-0.01$, i.e., $5.5\pm0.2$ standard deviations away from the local bound.}
\label{fig4}
\end{figure}

\textit{Experimental steering inequality violation.}--- To rule out any LHS model for the observed assemblage, we tested it against steering inequalities. The optimal inequality and the associated set of operators $\{F_{a|\theta}^\textrm{opt}\}_{a,\theta}$ was defined using the SDP mentioned previously. Figure~\ref{fig4}~(a) shows the Wigner functions of some experimental conditional states and the corresponding optimal operators. Their structure can be understood to some extent as $\mathcal{S}$ is found by integrating over the phase space the product of both functions and then summing over all $a$ and $\theta$. To obtain a negative value of $\mathcal{S}$, it is then apparent that the Wigner functions of the optimal operators should present high negativity in the area where the corresponding states exhibit greater Wigner function values. Applying these operators to our assemblage, we find a steering inequality violation $\mathcal{S}^{\textrm{opt}}\simeq-0.01$, in good agreement with our predicted value given the few percent uncertainty on the experimental losses.

The steering test can however only be successful after evaluating the associated error bar. As Alice is not trusted, only Bob's procedure has to be evaluated. The main source of error is the maximum likelihood reconstruction of the steered states. One currently used technique for quantifying tomographic uncertainties is bootstrapping \cite{Efron94}, which consists in resampling the data. While this procedure generally provides a good estimate, the results are not always reliable because the quantum state is inferred using a finite number of measurements and, moreover, the obtained bounds have no well-defined operational interpretation ~\cite{Christandl2012,blume-kohout_optimal_2010, Jung_increasing_2010, blume-kohout_robust_2012}. We therefore followed a different method, proposed in~\cite{faist_practical_2016} as \textit{quantum error bars} and based on robust confidence regions, to accurately evaluate the error bound in the EPR steering observation. While typical bootstrapping procedures rely only on the state most likely to have been observed, this method includes a broad exploration of the state space through a Metropolis-Hastings algorithm (see Supplemental Material for details and comparison of the two methods \cite{SM}). It therefore takes into account a wider range of states that are close to maximal likelihood but that may lead to a significantly different result in terms of steering violation. The resulting probability histogram of $\mathcal{S}$ is presented in Fig.~\ref{fig4}~(b). It shows that the steering inequality is violated with a separation of $5.5\pm0.2$ standard deviations from the LHS bound. This result represents a clear demonstration of EPR steering and a robust entanglement certification with one untrusted party. 

\textit{Conclusion.}--- In summary, we have presented a detailed study of EPR steering using for the first time a hybrid entangled state shared between two remote parties relying on different information encodings. Our implementation is based on two local homodyne setups: one to steer the CV system via quadrature measurements and sign binning on the DV side, the other one to perform quantum tomography of the resulting conditioned states. Because no post selection is used here, our steering test is free from detection loophole \cite{brunner_bell_2014,guerreiro_demonstration_2016,fuwa_experimental_2015}. An open question interesting to investigate will be whether one-way steering \cite{wiseman_steering_2007,Handchen2012,Bowles2014,Qin2017,Tischler2018}, i.e., steering that can occur from one party to the other and not the other way round, can be demonstrated with this class of hydrid states. Besides fully optical implementations of heterogeneous quantum networks, the entanglement certification presented here may also find extensions to a variety of developing hybridized platforms where CV oscillators are coupled to DV systems \cite{andersen_hybrid_2015}, such as microwave fields or mechanical oscillators coupled to two-level atom-like systems. \\

\begin{acknowledgments}
The authors thank O. Morin and K. Huang for their contributions in the early stage of the experiment. This work was supported by the European Research Council (Starting Grant HybridNet), Sorbonne Universit\'e (PERSU program) and the French National Research Agency (Hy-Light project). Part of this research was carried out at the Jet Propulsion Laboratory, California Institute of Technology, under a contract with the National Aeronautics and Space Administration. V. B. V. and S. W. N. acknowledge partial funding for detector development from the DARPA Information in a Photon (InPho) and QUINESS programs. G.G. is supported by the EU (Marie Curie fellowship).
\end{acknowledgments}


\section{References}

\begin{thebibliography}{7}

	
\bibitem{cavalcanti_quantum_2017} D.~Cavalcanti and P.~Skrzypczyk, Quantum steering: a review with focus on semidefinite programming, Rep. Prog. Phys. \textbf{80}, 024001 (2017).

\bibitem{wiseman_steering_2007} H.~M. Wiseman, S.~J. Jones, and A.~C. Doherty, Steering, Entanglement, Nonlocality, and the Einstein-Podolsky-Rosen Paradox, Phys. Rev. Lett. \textbf{98}, 140402 (2007).

\bibitem{brunner_bell_2014} N.~Brunner, D.~Cavalcanti, S.~Pironio, V.~Scarani, and S.~Wehner, Bell Nonlocality, Rev. Mod. Phys. \textbf{86}, 419 (2014).

\bibitem{branciard_one-sided_2012} C.~Branciard, E.~G. Cavalcanti, S.~P. Walborn, V.~Scarani, and H.~M. Wiseman, One-sided device-independent quantum key distribution: Security, feasibility, and the connection with steering, Phys. Rev. A \textbf{85}, 010301 (2012).

\bibitem{law_quantum_2014} Y.~Z. Law, L.~P. Thinh, J.-D. Bancal, and V.~Scarani, Quantum randomness extraction for various levels of characterization of the devices, J. Phys. A \textbf{47}, 424028 (2014).

\bibitem{passaro_optimal_2015} E.~Passaro, D.~Cavalcanti, P.~Skrzypczyk, and A.~Ac{\'i}n, Optimal randomness certification in the quantum steering and prepare-and-measure scenarios, New J. Phys. \textbf{17}, 113010 (2015).

\bibitem{Toth+2015EntSteering} G.~T\'{o}th, T.~Moroder, and O.~G\"{u}hne, Evaluating Convex Roof Entanglement Measures, Phys. Rev. Lett. \textbf{114}, 160501 (2015).

\bibitem{Cavalcalnti + Ass q networks} D.~Cavalcanti, P.~Skrzypczyk, G.~H.~Aguilar, R.~V.~Nery, P.~H.~Souto~Ribeiro, and S.~P.~Walborn, Detection of entanglement in asymmetric quantum networks and multipartite quantum steering, Nat. Commun. \textbf{6}, 7941 (2015).

\bibitem{McCutcheon} W.~McCutcheon \textit{et al.}, Experimental verification of multipartite entanglement in quantum networks, Nat. Commun. \textbf{7}, 13251 (2016).

\bibitem{Gheorghiu+Rigidity steering delegated qc} A.~Gheorghiu, P.~Wallden, and E.~Kashefi, Rigidity of quantum steering and one-sided device-independent verifiable quantum computation, New J. Phys. \textbf{19}, 023043 (2017).

\bibitem{SupicHoban16 self testing through steering} I.~\ifmmode \check{S}\else \v{S}\fi{}upi\ifmmode \acute{c}\else \'{c}\fi{} and M.~J.~Hoban, Self-testing through EPR-steering, New J. Phys. \textbf{18}, 075006 (2016).
	
\bibitem{saunders_experimental_2010} D.~J.~Saunders, S.~J.~Jones, H.~M.~Wiseman, and G.~J.~Pryde, Experimental {EPR}-steering using {Bell}-local states, Nat. Phys. \textbf{6}, 845 (2010).

\bibitem{smith_conclusive_2012} D.~H.~Smith \textit{et al.}, Conclusive quantum steering with superconducting transition-edge sensors, Nat. Commun. \textbf{3}, 625 (2012).

\bibitem{wittmann_loophole-free_2012} B.~Wittmann, S.~Ramelow, F.~Steinlechner,  N.~K.~Langford, N.~Brunner, H.~M.~Wiseman, R.~Ursin, and A.~Zeilinger, Loophole-free {Einstein}{\textendash}{Podolsky}{\textendash}{Rosen} experiment via quantum steering, New J. Phys. \textbf{14}, 053030 (2012).

\bibitem{weston_heralded_2018} M.~M.~Weston, S.~Slussarenko, H.~M.~Chrzanowski, S.~Wollmann, L.~K.~Shalm, V.~B.~Verma, M.~S~Allman, S.~W.~Nam, and G.~J.~Pryde, Heralded quantum steering over a high-loss channel, Sci. Adv. \textbf{4},  e1701230 (2018).


\bibitem{fuwa_experimental_2015} M.~Fuwa, S.~Takeda, M.~Zwierz, H.~M. Wiseman, and A.~Furusawa, Experimental proof of nonlocal wavefunction collapse for a single particle using homodyne measurements, Nat. Commun. \textbf{6}, 6665 (2015).

\bibitem{guerreiro_demonstration_2016} T.~Guerreiro \textit{et al.}, Demonstration of {Einstein}-{Podolsky}-{Rosen} {Steering} {Using} {Single}-{Photon} {Path} {Entanglement} and {Displacement}-{Based} {Detection}, Phys. Rev. Lett. \textbf{117}, 070404 (2016).


\bibitem{Handchen2012} V. H\"andchen, T. Eberle, S. Steinlechner, A. Samblowski, T. Franz, R. F. Werner, and R. Schnabel, Observation of one-way Einstein-Podolsky-Rosen steering, Nat. Photonics \textbf{6}, 596 (2012).
\bibitem{Armstrong2015} S. Armstrong \textit{et al.}, Multipartite Einstein-Podolsky-Rosen steering and genuine tripartite entanglement with optical networks, Nat. Phys. \textbf{11}, 167 (2015).
\bibitem{walk_experimental_2016} N.~Walk \textit{et al.}, Experimental demonstration of Gaussian protocols for one-sided device-independent quantum key distribution, Optica \textbf{3}, 634 (2016).
\bibitem{Deng2017} X. Deng \textit{et al.}, Demonstration of Monogamy Relations for Einstein-Podolsky-Rosen Steering in Gaussian Cluster States, Phys. Rev. Lett. \textbf{118}, 230501 (2017).
\bibitem{Qin2017} Z. Qin, X. Deng, C. Tian, M. Wang, X. Su, C. Xie, and K. Peng, Manipulating the direction of Einstein-Podolsky-Rosen steering, Phys. Rev. A \textbf{95}, 052114 (2017).


\bibitem{van_loock_optical_2011} P.~van Loock, Optical Hybrid Approaches to Quantum Information, Laser Photonics Rev. \textbf{5}, 167 (2011).

\bibitem{andersen_hybrid_2015} U.~L. Andersen, J.~S. Neergaard-Nielsen, P.~van Loock, and A.~Furusawa, Hybrid discrete- and continuous-variable quantum information, Nat. Phys. \textbf{11}, 713 (2015).

\bibitem{takeda_deterministic_2013} S.~Takeda, T.~Mizuta, M.~Fuwa,  P.~van Loock, and A.~Furusawa, Deterministic quantum teleportation of photonic quantum bits by a hybrid technique, Nature (London) \textbf{500}, 315 (2013).

\bibitem{morin_witnessing_2013} O.~Morin, J.-D.~Bancal, M.~Ho, P.~Sekatski, V.~D'Auria, N.~Gisin, J.~Laurat, and N.~Sangouard, Witnessing {Trustworthy} {Single}-{Photon} {Entanglement} with {Local} {Homodyne} {Measurements}, Phys. Rev. Lett. \textbf{110}, 130401 (2013).

\bibitem{ho_witnessing_2014} M.~Ho, O.~Morin, J.-D.~Bancal, N.~Gisin, N.~Sangouard, and J.~Laurat, Witnessing single-photon entanglement with local homodyne measurements: Analytical bounds and robustness to losses, New J. Phys. \textbf{16}, 103035 (2014).

\bibitem{jeong_generation_2014} H.~Jeong, A.~Zavatta, M.~Kang, S.-W.~Lee, L.~S. Costanzo, S.~Grandi, T.~C. Ralph, and M.~Bellini, Generation of hybrid entanglement of light, Nat. Photonics \textbf{8}, 564 (2014).

\bibitem{morin_remote_2014} O.~Morin, K.~Huang, J.~Liu, H.~Le~Jeannic, C.~Fabre, and J.~Laurat, Remote creation of hybrid entanglement between particle-like and wave-like optical qubits, Nat. Photonics \textbf{8}, 570 (2014).

\bibitem{RSP} H. Le Jeannic, A. Cavaill\`{e}s, J. Raskop, K. Huang, and J. Laurat, Remote preparation of continuous-variable qubits using loss-tolerant hybrid entanglement of light, Optica \textbf{5}, 1012 (2018).

\bibitem{SM} See Supplemental Material for derivation of the conditional states, and for error bar computation and comparison of different methods. It includes Ref. \cite{SM1} and \cite{SM2}.
\bibitem{SM1} G.~O.~Roberts and J.~S.~Rosenthal, Optimal Scaling for Various {Metropolis}-{Hastings} Algorithms, Stat. Sci. \textbf{16}, 351 (2001).
\bibitem{SM2} V.~Ambegaokar and M.~Troyer, Estimating errors reliably in {Monte} {Carlo} simulations of the {Ehrenfest} model, Am. J. Phys. \textbf{78}, 150 (2010).

\bibitem{vandenberghe_semidefinite_1996} L.~Vandenberghe and S.~Boyd, Semidefinite {Programming}, SIAM Rev. \textbf{38}, 49 (1996).

\bibitem{jeannic_high-efficiency_2016} H.~Le Jeannic, V.~B. Verma, A.~Cavaill{\`e}s, F.~Marsili, M.~D. Shaw, K.~Huang, O.~Morin, S.~W. Nam, and J.~Laurat, High-efficiency {WSi} superconducting nanowire single-photon detectors for quantum state engineering in the near infrared, Opt. Lett. \textbf{41}, 5341 (2016).

\bibitem{morin_quantum_2014} O.~Morin, J.~Liu, K.~Huang, F.~Barbosa, C. Fabre, and J.~Laurat, Quantum state engineering of light with continuous-wave optical parametric oscillators, J. Vis. Exp. \textbf{87}, e51224 (2014).


\bibitem{lvovsky_iterative_2004} A. I.~Lvovsky, Iterative maximum-likelihood reconstruction in quantum homodyne tomography, J. Opt. B \textbf{6}, S556 (2004).

\bibitem{morin_experimentally_2013} O.~Morin, C.~Fabre, and J.~Laurat, Experimentally {Accessing} the {Optimal} {Temporal} {Mode} of {Traveling} {Quantum} {Light} {States}, Phys. Rev. Lett. \textbf{111}, 213602 (2013).

\bibitem{huang_microcontroller-based_2014} K.~Huang, H.~Le~Jeannic, J.~Ruaudel, O.~Morin, and J.~Laurat, Microcontroller-based locking in optics experiments, Rev. Sci. Instrum. \textbf{85}, 123112 (2014).

\bibitem{Vidal} G. Vidal and R. F. Werner, Computable measure of entanglement, Phys. Rev. A \textbf{65}, 032314 (2002).

\bibitem{Efron94} B. Efron and R. J. Tibshirani, An Introduction to the Bootstrap (CRC press, 1994).

\bibitem{Christandl2012} M. Christandl and R. Renner, Reliable Quantum State Tomography, Phys. Rev. Lett. \textbf{109}, 120403 (2012).

\bibitem{blume-kohout_optimal_2010} R.~Blume-Kohout, Optimal, reliable estimation of quantum states, New J. Phys. \textbf{12}, 043034 (2010).

\bibitem{Jung_increasing_2010} B.~Jungnitsch, S.~Niekamp, M.~Kleinmann, O.~G\"uhne, H.~Lu, W.-B.~Gao, Y.-A.~Chen, Z.-B.~Chen, and J.-W. Pan, Increasing the Statistical Significance of Entanglement Detection in Experiments, Phys. Rev. Lett. \textbf{104}, 210401 (2010).
	
\bibitem{blume-kohout_robust_2012} R.~Blume-Kohout, Robust error bars for quantum tomography, eprint arXiv:1202.5270.

\bibitem{faist_practical_2016} P.~Faist and R.~Renner, Practical and Reliable Error Bars in Quantum Tomography, Phys. Rev. Lett. \textbf{117}, 010404 (2016).

\bibitem{Bowles2014} J. Bowles, T. V\'ertesi, M. T. Quintino, and N. Brunner, One-way Einstein-Podolsky-Rosen Steering, Phys. Rev. Lett. \textbf{112}, 200402 (2014). 
\bibitem{Tischler2018} N. Tischler \textit{et al.}, Conclusive Experimental Demonstration of One-Way Einstein-Podolsky-Rosen Steering, Phys. Rev. Lett. \textbf{121}, 100401 (2018).


\end{thebibliography}
\end{document}